\documentclass[a4paper,11pt]{article}
\pdfoutput=1
\evensidemargin=\oddsidemargin
\usepackage[T1]{fontenc}
\usepackage{pstricks}
\usepackage{color}
\usepackage{slashed}
\usepackage{graphicx,graphics}
\usepackage{verbatim}
\usepackage{amssymb}
\usepackage{amsthm}
\usepackage{array}
\usepackage{enumerate}
\usepackage{amsmath,amsfonts}
\usepackage{caption}
\usepackage{subcaption}
\usepackage[colorlinks=true,
            linkcolor=red,
            urlcolor=blue,
            citecolor=blue] {hyperref}
\usepackage{cite}

\def\ls{\mathrel{\lower4pt\vbox{\lineskip=0pt\baselineskip=0pt
           \hbox{$<$}\hbox{$\sim$}}}}
\def\gs{\mathrel{\lower4pt\vbox{\lineskip=0pt\baselineskip=0pt
           \hbox{$>$}\hbox{$\sim$}}}}

\allowdisplaybreaks
\begin{document}

\begin{flushright}
        {\small MI-TH-1517}\\
        {\small CETUP2015-006}
\end{flushright}

\begin{center}
        {\Large \bf Distinguishing Neutrino Mass Hierarchies using Dark Matter Annihilation Signals at IceCube} \\
        \vspace*{1cm} {\sf Rouzbeh Allahverdi
        $^{a,}$\footnote{rouzbeh@unm.edu},~Bhaskar Dutta$^{b,}$\footnote{dutta@physics.tamu.edu}, ~Dilip Kumar Ghosh$^{c,}$\footnote{tpdkg@iacs.res.in},
        ~Bradley Knockel$^{a,}$\footnote{knockel.physics@gmail.com}, ~Ipsita Saha$^{c,}$\footnote{tpis@iacs.res.in}} \\
        \vspace{10pt} {\small } {\em $^a$ Department of Physics and Astronomy, University of New Mexico, Albuquerque, NM 87131, USA. \\
                $^b$ Department of Physics and Astronomy, Mitchell Institute for Fundamental Physics and Astronomy,
                     Texas A $\&$ M University, College Station, TX 77843-4242, USA.\\
                $^c$ Department of Theoretical Physics,
                Indian Association for the Cultivation of Science, \\
                2A $\&$ 2B, Raja S.C. Mullick Road, Kolkata 700032, India.}

        \normalsize
\end{center}

\begin{abstract}
We explore the possibility of distinguishing neutrino mass hierarchies through the neutrino 
signal from dark matter annihilation at neutrino telescopes.
We consider a simple extension of the standard model 
where the neutrino masses and mixing angles are obtained via the type-II seesaw mechanism as an explicit example. We
show that future extensions of IceCube neutrino telescope may detect the neutrino signal from DM annihilation at the Galactic Center and inside the Sun, and differentiate between the normal and inverted mass hierarchies, in this model.
\end{abstract}

\bigskip
%%%%%%%%%%%%%%%%%%%%%%%%%%%%%%%%%%%%%%%%%%%%%%%%%%%%%%%%%%
\section{Introduction}
Although various lines of evidence support the existence of dark matter (DM) in the universe,
its identity remains as a mystery.
Among various proposed candidates, the weakly interacting massive particles (WIMPs) are 
promising ones that typically arises in many extensions of the standard model (SM)~\cite{Jungman:1995df,Bertone:2004pz}. 
Major direct, indirect, and collider experimental searches are currently underway 
to detect the particle nature of DM and determine its properties.

Just like DM, the origin of neutrino mass
and mixing defines a very interesting area of investigation beyond the SM (BSM)~\cite{Strumia:2006db,King:2014nza}.
These are the two encouraging avenues of new physics and
a large number of BSM scenarios have already been proposed to address
them. It will be even more interesting to investigate models where the two sectors are connected.
In fact, such a connection arises naturally in a class of models where 
DM is tied to the neutrino sector. For example, in type-II seesaw plus a 
singlet scalar scenario \cite{Gogoladze:2009gi}, it has been shown
that the neutrino mass hierarchy can influence the
DM annihilation rate to charged leptons as well as neutrinos and
subsequently provide a possible explanation of the positron fraction excess
observed at the AMS-02 experiment~\cite{Dev:2013hka}. Similar
studies have been done in the $B-L$ extension of the minimal supersymmetric standard model (MSSM)
where it is shown that the annihilation of the right-handed sneutrino DM produces 
neutrino final states with a distinct feature~\cite{Allahverdi:2014eca}.
There have been studies on the possibility of discovering the $B-L$ sneutrino DM at the LHC~\cite{Belanger:2011ny,BhupalDev:2012ru,Arina:2013zca}. The connection between the DM and neutrinos
in such scenarios can also be probed through DM direct detection~\cite{An:2011uq} and some indirect detection experiments~\cite{Ghosh:2014pwa}.
Therefore, a combination of direct, indirect and collider signatures could be useful to
explore this connection (for a review, see~\cite{Lattanzi:2014mia}).

Our focus here is on the DM indirect detection signals as a means of testing
the connection between the DM and neutrinos. Neutrino telescopes like IceCube are able to examine these 
models through the signal arising from the DM annihilation into neutrinos at 
the Galactic Center~\cite{Aartsen:2012kia} and inside the Sun~\cite{Aartsen:2015xej}. 
%Also, DM indirect detection searches 
%like the Fermi-Lat experiment~\cite{Albert:2015zma} can probe these models via
%the photon signal coming from the DM annihilation at the Galactic Center.
Unlike the photon flux, 
the neutrino flux arising from the DM annihilation
at the Galactic Center has less astrophysical background uncertainty, which
allows us to probe the exact nature of the DM more accurately. 
However, the model prediction for the photon fluxes coming from the DM annihilation 
at the Galactic Center should be consistent with the Fermi-LAT~\cite{Albert:2015zma} data on the gamma-ray flux.
Therefore, it will be interesting to combine the results from IceCube and Fermi-LAT to study
models where the DM and the neutrino sectors are interconnected.

In this paper, we study these issues within the above mentioned extension of the 
SM~\cite{Gogoladze:2009gi,Dev:2013hka}. In this model, the light neutrino masses and 
mixing angles are generated using the well-known type-II seesaw mechanism~\cite{Schechter:1980gr,Magg:1980ut,Cheng:1980qt,Mohapatra:1980yp,Lazarides:1980nt}
that introduces $SU(2)_L$ triplet scalars.
The neutral component of the triplet acquires a non-zero vacuum expectation value (vev) that
generates tiny neutrino masses by breaking lepton number by two
units and mixing among different neutrino flavors.
Among the SM particles, these triplet scalars only couple to gauge bosons
through gauge couplings and to leptons via Yukawa couplings. These Yukawa
couplings are related to the triplet vev, which in turn determines
different decay modes of these triplet scalars.
It has been observed~\cite{Perez:2008ha,Melfo:2011nx} that the triplets will dominantly
decay to a pair of leptons if the vev is less than 0.1 MeV,
otherwise the gauge boson final state will become dominant. In order
to accommodate a stable DM candidate, the model is augmented by a SM singlet
scalar with a $Z_2$ symmetry~\cite{Gogoladze:2009gi}.
In the minimal case, the DM only couples to the Higgs and to the triplet scalars.
For a vev smaller than 0.1 MeV, the annihilation of the DM produces a pair 
of triplets that will further decay to
SM leptons (including neutrinos) with almost $100\%$ branching fraction. 
Therefore, the flavor composition of the final states will depend
upon the neutrino mass hierarchy that sets the Yukawa couplings. We exploit this feature
and show how different neutrino mass hierarchies can be distinguished by 
using the muon signal arising from conversion of muon neutrinos from 
DM annihilation at the Galactic Center and inside the Sun at IceCube.
We note that the photon signal from the Galactic Center does not 
discriminate amongst different hierarchies because the difference will not  
be significant.

The paper is organized as follows. In Sec.~\ref{model}, we discuss the model in brief. In Sec.~\ref{analysis},
we present the analysis of the neutrino signal from
DM annihilation at the Galactic Center and inside the Sun for normal and inverted hierarchies.
We also calculate the corresponding muon spectra at the detector and compare them with the background.
Finally, in Sec.~\ref{conclusion} we present our discussion and conclusion.
%%%%%%%%%%%%%%%%%%%%%%%%%%%%%%%%%%%%%%%%%%%%%%%%%%%%%
\section{Type-II seesaw with a scalar singlet DM candidate}\label{model}
The minimal version of the type-II seesaw model is an extension of the SM
that includes a $SU(2)_L$ scalar triplet $\Delta$ with hypercharge $Y=2$.
\begin{eqnarray}
\mathbf{\Delta} = \left(\begin{array}{cc}
\frac{\Delta^+}{\sqrt{2}} & \Delta^{++} \\
\Delta^0 & \frac{\Delta^-}{\sqrt{2}}
\end{array}\right)
\end{eqnarray}
In order to accommodate a cold dark matter (CDM) candidate, we introduce a SM singlet real scalar $D$.
The stability of the DM candidate is ensured by imposing a $Z_2$ symmetry under which $D$ is
odd while all other SM particles and the triplet $\Delta$ are even.

The scalar potential of the model, including the relevant terms for the DM candidate $D$, is given by~\cite{Dev:2013hka}
\begin{eqnarray}
{\cal V}(\Phi,\mathbf{\Delta}) &=& -m_\Phi^2(\Phi^\dag \Phi)+\frac{\lambda}{2}(\Phi^\dag \Phi)^2+M^2_\Delta
{\rm Tr}(\mathbf{\Delta} ^\dag \mathbf{\Delta})+ \frac{\lambda_1}{2}\left[{\rm Tr}(\mathbf{\Delta} ^\dag \mathbf{\Delta})\right]^2\nonumber \\
&& +\frac{\lambda_2}{2}\left(\left[{\rm Tr}(\mathbf{\Delta} ^\dag \mathbf{\Delta})\right]^2-{\rm Tr}
\left[(\mathbf{\Delta} ^\dag \mathbf{\Delta})^2\right]\right)+\lambda_4(\Phi^\dag \Phi){\rm Tr}(\mathbf{\Delta} ^\dag \mathbf{\Delta}) \nonumber \\
&&+\lambda_5\Phi^\dag[\mathbf{\Delta}^\dag,\mathbf{\Delta}]\Phi+\left(\frac{\Lambda_6}{\sqrt 2}\Phi^{\sf T}i\sigma_2\mathbf{\Delta}^\dag \Phi+{\rm h.c.}\right)\,,
\label{scalar_potential}\\
{\cal V}_{\rm DM}(\Phi,\mathbf{\Delta}, D) &=& \frac{1}{2}m_D^2D^2+
\lambda_D D^4+\lambda_\Phi D^2(\Phi^\dag \Phi)+
\lambda_\Delta D^2{\rm Tr}(\mathbf{\Delta}^\dag \mathbf{\Delta})\,.
\label{dm_potential}
\end{eqnarray}
where $\Phi$ is the SM Higgs doublet. The couplings $\lambda_i ( i = 1-5) $
can be taken as real without any loss of generality. Due to spontaneous
symmetry breaking, the non-zero vev of the Higgs
doublet generates a tadpole term through interaction involving the 
$\Lambda_6$ coupling in Eq.~(\ref{scalar_potential}). This in turn induces a
non-zero vev ($v_\Delta$) for the neutral component of the triplet thereby breaking
lepton number by two units. This triplet vev contributes to the SM 
gauge boson masses at the tree level which leads to a deviation in the 
electroweak $\rho$ parameter. Hence, $v_\Delta$ is constrained by the
electroweak precision data that requires the $\rho$ parameter to be close 
to the SM value of unity \cite{pdg2014} which eventually demands
$v_\Delta < 2 ~\rm GeV$ \cite{delAguila:2008ks}. In addition to this, 
current experimental bounds on lepton flavor violating processes
put a lower bound on $v_\Delta$ \cite{Akeroyd:2009nu,Fukuyama:2009xk,Kanemura:2012rs} as
\begin{eqnarray}
v_\Delta M_\Delta \geq 150 ~\rm eV~ GeV \,.
\end{eqnarray}
It is to be mentioned here that we assume negligible mixing between the
doublet and triplet scalars in our analysis \footnote{Detailed expressions for
physical scalar mass eigenstates can be found in \cite{Dev:2013ff}.}
and with this assumption, the DM potential of Eq.~(\ref{dm_potential}) can be expressed in terms of physical scalars
$(h, H^\pm, H^{\pm\pm},H^0,A^0)$ as
\begin{eqnarray}
{\cal V}_{\rm DM}(\Phi,\mathbf{\Delta}, D) &=& \frac{1}{2}m_{DM}^2D^2+
\lambda_D D^4+\lambda_\Phi v D^2 h+\frac{1}{2}\lambda_\Phi D^2 h^2 + \nonumber \\
&&\lambda_\Delta D^2\left[H^{++}H^{--} + H^+H^- + \frac{1}{2}\left(H_0^2 +A_0^2\right) + \frac{1}{2} v_\Delta H_0 \right]\,.
\label{dm_mass_potential}
\end{eqnarray}
Here, $m_{DM}^2 = m_D^2 + \lambda_\phi v^2 + \lambda_\Delta v_\Delta^2$ denotes the physical mass of the DM candidate.
It is evident from Eq.~(\ref{dm_mass_potential}) that the DM candidate can annihilate to a pair of Higgs and to a pair of
exotic triplet scalar particles through the coupling parameters $\lambda_\phi$ and $\lambda_\Delta$. In the limit
$m_{DM} > m_\Delta$, where, $m_\Delta$ is the degenerate mass of all triplet scalars, the annihilation cross section
of the DM candidate (non-relativistic) is expressed as
\begin{eqnarray}
\left<\sigma v\right> &=& \frac{1}{16 \pi m_{DM}^2} \left[\lambda_\phi^2 \left(1 - \frac{m_h^2}{m_{DM}^2} \right) + 6 \lambda_\Delta^2 \left(1 - \frac{m_\Delta^2}{m_{DM}^2}\right)  \right]\,.
\label{anni_xsec}
\end{eqnarray}
This should yield the correct relic abundance of the DM particle that lies within the $3\sigma$ limit of current Planck bound
$\Omega_{DM}h^2 = 0.1199\pm0.0027$ \cite{Ade:2013zuv}. Now, the final state of DM annihilation will depend upon the
branching fraction of the Higgs and the triplet scalar decay. The triplet scalars can couple to the SM gauge bosons
through gauge coupling and to SM lepton doublets through Yukawa couplings.
\begin{eqnarray}
{\cal L}_Y &=& {\cal L}_Y^{\rm SM} -\frac{1}{\sqrt 2}\left (Y_\Delta\right)_{ij} L_i^{\sf T}Ci\sigma_2\Delta L_j
 +{\rm h.c.}
\end{eqnarray}
Here, $C$ is the charge conjugation operator and $L_i = (\nu_i, \ell_i)^T_L$ is the $SU(2)_L$ lepton doublet
with $i$ being the three generation indices.
Further, these Yukawa couplings can be obtained from the Majorana mass matrix of neutrinos that arises due to the non-zero
triplet vev $v_\Delta$.
\begin{eqnarray}
(M_\nu)_{ij} &=& v_\Delta(Y_\Delta)_{ij}\,, \\
Y_\Delta &=& \frac{M_\nu}{v_\Delta} = \frac{1}{v_\Delta}U^{\sf T}M_\nu^{\rm diag}U\,,
\label{YD}
\end{eqnarray}
where $M_\nu^{\rm diag} = {\rm diag}(m_1,m_2,m_3)$ and $U$ is the PMNS mixing matrix.

With the recent global analysis data (see Eq.~(\ref{neu_osc_data}) in the Appendix), and after using Eq.~(\ref{YD}),
we obtain the following structure of the Yukawa coupling matrix for the normal hierarchy (NH) and inverted hierarchy (IH) scenarios
\begin{comment}
\begin{eqnarray}
Y_\Delta &=& \frac{10^{-2}~{\rm eV}}{v_\Delta}\times \left\{\begin{array}{cc}
\left(\begin{array}{ccc}
0.88-0.30i & -1.29+0.09i & 1.25-0.37i \\
-1.29+0.09i & 1.72+0.25i & -1.58-0.25i \\
1.25-0.37i & -1.58-0.25i & 3.10+0.0i
\end{array}\right) & ({\rm normal~hierarchy}) \label{nh_yuk} \\
\left(\begin{array}{ccc}
4.02+0.30i &  0.94 - 0.15 i & -1.47 - 0.83 i \\
 0.94 - 0.15 i &  3.51 - 0.34 i & -1.95 - 0.56 i \\
-1.47 - 0.83 i & -1.95 - 0.56 i &
2.03 + 0.07 i \end{array}\right) & ({\rm inverted~hierarchy})\,. \label{ih_yuk}\\ 
\end{array}\right.
\label{yukawa}
\end{eqnarray}
\end{comment}
\begin{eqnarray}
Y_\Delta &=& \frac{10^{-2}~{\rm eV}}{v_\Delta}\times \left\{\begin{array}{cc}
\left(\begin{array}{ccc}
1.08-0.29i & -1.55+0.09i & 1.23-0.31i \\
-1.55+0.09i & 2.07+0.26i & -1.59-0.21i \\
1.23-0.31i & -1.59-0.21i & 2.59+0.0i
\end{array}\right) & ({\rm normal~hierarchy}) \label{nh_yuk} \\
\left(\begin{array}{ccc}
3.84+0.34i &  1.21 - 0.13 i & -1.39 - 0.94 i \\
1.21 - 0.13 i &  2.97 - 0.35 i & 1.98 - 0.65 i \\
-1.39 - 0.94 i & 1.98 - 0.65 i &
2.66 + 0.01 i \end{array}\right) & ({\rm inverted~hierarchy})\,. \label{ih_yuk}\\ 
\end{array}\right.
\label{yukawa}
\end{eqnarray}
%For Normal (NH) and Inverted (IH) hierarchy, we choose the lightest neutrino mass eigenvalue
%to be zero to determine Eq.~(\ref{yukawa}). 
To find the Yukawa matrix for the degenerate case,  we impose the $95\%$ C.L. limit  
on the sum of all light neutrino masses $\sum_{i} m_i < 0.23 ~\rm eV$ from Planck~\cite{Ade:2013zuv}, which 
uses the Baryon Acoustic Oscillation (BAO) and Cosmic Microwave Background (CMB) data.
By choosing $m_1 \simeq m_2 \simeq m_3 = 0.07 ~\rm eV$, we find
\begin{eqnarray}
Y_\Delta &=&\frac{10^{-2}~{\rm eV}}{v_\Delta}\times \begin{array}{cc}
\left(\begin{array}{ccc}
6.80-0.06i &  -0.13 - 0.04 i & 0. - 1.65 i \\
-0.13 - 0.04 i &  6.91 - 0.03 i & 0. - 1.10 i \\
0. - 1.65 i & 0. - 1.10 i &
6.71 + 0.10 i \end{array}\right) & ({\rm degenerate ~case})\,. \label{degen_yuk} 
\end{array}
%\label{yukawa}
\end{eqnarray}
%To get the Yukawa coupling for the degenerate case of Eq.~(\ref{degen_yuk}), we impose the $95\%$ C.L. limit  
%on the sum of all light neutrino masses $\sum_{i} m_i < 0.23 ~\rm eV$ from Planck~\cite{Ade:2013zuv}
%using the Baryon Acoustic Oscillation (BAO) and Cosmic Microwave Background (CMB) data.
%We choose $m_1 \simeq m_2 \simeq m_3 = 0.07 ~\rm eV$. 
However, as we will see later, the degenerate scenario does
not offer any significant information in our study.

The triplet scalar decay to leptonic final states,
$H^{++} \to \ell^+\ell^+ \,, H^+ \to \ell^+ \nu_\ell \,, H_0/A_0 \to \nu \nu $
with almost $100\%$ branching ratio when $v_\Delta \leq 0.1 ~\rm MeV$.
On the other hand, for larger $v_\Delta (> 0.1 ~\rm MeV)$,
gauge boson final states become dominant with the
decay modes $H^{++} \to W^+W^+ \,, H^+ \to W^+ Z \,, H_0/A_0 \to Z Z/W^+ W^-$~\footnote{Throughout this analysis we assume degenerate triplet scalars.}.
For almost same order
 of $\lambda_\phi$ and $\lambda_\Delta$ within the range for producing correct relic density,
 the DM can annihilate to three different final states (depending upon the branching ratios i.e on $v_\Delta$ for triplets) :
 \begin{enumerate}[(i)]
 \item neutrino and charged lepton final states.
 \item $W^\pm W^\mp$, $ZZ$ finals states
mainly from triplet scalar decays;
 $W W^*$, $ZZ^*$ finals states arise from the decay of the SM Higgs boson.
 \item $b\,,\tau$ final states from the decay of the SM Higgs boson.
 \end{enumerate}
We choose two extreme values of $v_\Delta$, namely, 1 eV and 1 GeV,
respectively to incorporate the effect of triplet scalars
decay to only leptonic final state or only to non-leptonic final states. Our study is mainly focused on the
study of neutrino flux coming from DM annihilation which can be obtained from all the above three final states.
However, for triplet scalars decaying into leptonic final states includes the contribution of Yukawa couplings $Y_\Delta$
which is a function of neutrino mass hierarchy and we expect to see the difference in the neutrino flux
between the NH and IH scenarios.
%different hierarchies, say Normal (NH) or Inverted (IH). 
In the following sections, we will explicitly
show how our analysis depends upon the neutrino mass hierarchy.

In this regard, it is to be noted that the
triplet scalars can produce interesting signals at the colliders.
The strongest limit on the scalar masses comes from the current
searches at the LHC for the signature of doubly charged scalar where
they can be pair produced via Drell-Yan and Vector Boson fusion
and then decay to a pair of same-sign leptons. 
%The experimental
%lower bound on $m_{H^{\pm \pm}}$ has been set by the
%CMS experiment~\cite{Chatrchyan:2012ya} from
%3 isolated lepton search for $(H^{\pm\pm} \to e^\pm e^\pm)$ and
%$(H^{\pm\pm} \to \mu^\pm \mu^\pm)$ decay modes at the center of mass energy 7 TeV. The lower mass limits 
%are respectively $383$ GeV and $408$ GeV, assuming
%$100\%$ branching ratio to individual channels.
%The 7 TeV result of ATLAS searches for the doubly charged scalar
%decaying to same-sign dimuon channel excludes $m_{H^{\pm \pm}}$ below $355$ GeV~\cite{Aad:2012xsa}
%for 3 muon final state and the ATLAS bound~\cite{ATLAS:2012hi} coming from 4 lepton final state search 
%demands $m_{H^{\pm \pm}} > 409$ GeV and $m_{H^{\pm \pm}} > 398$ GeV 
%for the corresponding decay of $H^{\pm\pm}$ to $e^\pm e^\pm$ and
%$\mu^\pm \mu^\pm$ only. 
The experimental
lower bound on $m_{H^{\pm \pm}}$ has been set by the
ATLAS experiment~\cite{ATLAS:2014kca} from
a pair of isolated lepton search for $(H^{\pm\pm} \to e^\pm e^\pm)$ and
$(H^{\pm\pm} \to \mu^\pm \mu^\pm)$ decay modes at the center of mass energy 8 TeV.
The 95\% C.L. lower limit on the doubly charged Higgs mass for the same-sign
isolated muons final state exclude mass range below 516 GeV. However, the experimental
lower limit is based on the assumption that the doubly charged scalar decay to the dimoun channel
with 100\% branching ratio which is not the case in our scenario. In this model, the doubly charged scalar
decay to dimuon channel with atmost 30\% branching ratio and thus the lower limit on $m_{H^{\pm \pm}}$
can be relaxed. Following Fig. 5 of Reference~\cite{ATLAS:2014kca},
%In order to satisfy the
%above experimental results, 
we set the degenerate mass of the
triplets ($m_{\Delta}$) as 400 GeV throughout our analysis.
%%%%%%%%%%%%%%%%%%%%%%%%%%%%%%%%%%%%%%%%%%%%%%%%%%%%%%%%%%%%%%%%
\section{Analysis}\label{analysis}
In this section, we demonstrate our findings on the DM annihilation
at the Galactic Center and inside the Sun and the possibilities
of using the respective neutrino signals to distinguish the NH and IH scenarios. For the purpose of our analysis,
we choose to work with four benchmark points, shown in Table~\ref{bp},
by fixing $v_\Delta$ for two DM masses. We choose the DM mass
to be greater than 400 GeV to get on-shell triplet scalars
in the pair production processes of DM annihilation. Our choice
of $v_\Delta$ has already been justified in the previous section.
We choose
$\lambda_\Delta$ and $\lambda_\Phi$ couplings in the model (see Eq.~(\ref{dm_potential}))
such that the correct DM abundance is obtained via thermal freeze-out with nominal value for thermally averaged
annihilation cross-section $\left<\sigma_{\rm ann} v\right> \simeq 3\times 10^{-26}~{\rm cm^3/s}$.
$\lambda_\Phi$ also enters into the direct detection cross section $(\sigma_{SI})$
through the $t$-channel exchange of the SM Higgs. Values of $\lambda_\Phi$ that satisfy thermal relic abundance yield a direct
detection cross section well below the current limits from LUX experiment~\cite{Akerib:2013tjd}.
The choice of $\lambda_\Delta$ and $\lambda_\Phi$
and the corresponding relic density and direct detection cross section
are also listed in Table~\ref{bp}. The correct DM relic abundance can also be produced via
non-thermal mechanisms~\cite{Barrow:1982ei,McDonald:1989jd,Kamionkowski:1990ni} in which case a larger DM annihilation cross section is allowed. We will discuss the implications of a larger annihilaiton cross section at the end of this section.  

As mentioned earlier, the DM annihilation will give rise to neutrino
fluxes of different flavors that will help in distinguishing the structure
of Yukawa coupling and, hence, the neutrino mass hierarchy.
The IceCube detects neutrinos 
by recording the Cherenkov light from relativistic 
charged particles in its volume. Muon neutrinos ($\nu_\mu$)
produce muon ($\mu$) tracks via the charged current interactions in
the detector. On the other hand, electron neutrinos ($\nu_e$) and tau neutrinos ($\nu_\tau$) result in
hadronic and electromagnetic cascade events in the ice. Since the cascades
are localized, 
they carry no directional information, and hence
are not good enough for performing a meaningful DM search over the background.
For this reason, we focus on $\nu_\mu$ fluxes that arrive on
the Earth. We also estimate the photon flux for all scenarios in our model.
\begin{table}[htbp!]
	\centering
	\begin{tabular}{|c|c|c|c||c|c|c|c|}
		\hline
		Benchmarks & $v_\Delta$ & $m_{\rm DM}$(GeV)& $m_{\Delta}$(GeV) & $\lambda_\Delta$ & $\lambda_\phi$ & relic density & $\sigma_{SI}$ (pb)  \\ \hline
		BP1  & 1 eV & 500&400 & 0.055 & 0.04 & 0.117 & $2.25\times10^{-10}$\\  \hline
		BP2 & 1 eV & 700&400 & 0.075 & 0.05 & 0.122 & $1.79\times10^{-10}$\\ \hline
		BP3  & 1 GeV & 500&400 & 0.055 & 0.04 & 0.117 & $2.25\times10^{-10}$ \\  \hline
		BP4 & 1 GeV & 700&400 & 0.075 & 0.05 & 0.122 & $1.79\times10^{-10}$ \\\hline
		\hline
	\end{tabular}
	\caption{Benchmark Points (BPs) and the corresponding DM relic density and direct detection cross-sections. }
	\label{bp}
\end{table}
Cosmic ray showers create a muon background
that can be controlled by selecting only the upward going
events since muons are stopped in the Earth. This
limits the observation of DM signal to the time when the source
is below the horizon. With atmospheric muons thus eliminated, 
the most significant
contribution to the remaining background comes
from atmospheric neutrinos~\cite{Honda:2006qj}.
In addition to this, a portion of the detector may be
used as a veto to observe the contained muon events with a conversion
vertex inside the instrumented volume, as in the case of DeepCore
array in IceCube. The veto procedure virtually eliminates the contribution to the
background from atmospheric muons by selecting
only contained vertices. This increases the potential observation
time to the full year when the source is both above and below the
horizon.

We use both contained and through-going muons in our analysis. For contained
muon tracks, the vertex at which $\nu_\mu$ is converted to a
muon, is within the detector volume. Through-going muons 
represent those events that go through a surface inside the detector 
but may have been produced outside it. To acquire the muon spectra, we use three simple methods:
\begin{enumerate}
	\item WimpSim~\cite{Blennow:2007tw} automatically provides muons converted from solar neutrinos.
	\item DarkSUSY~\cite{Gondolo:2004sc} gives muons from atmospheric neutrinos (which we average to make an isotropic approximation).
	\item Finally, GENIE~\cite{Andreopoulos:2009rq} provides muons converted from neutrinos coming from the Galactic Center.
\end{enumerate}
GENIE provides the spectrum of contained muons produced from $\nu_\mu$s arriving at the detector.
To find the total number of muon events, we use the total deep-inelastic charged-current
cross sections from reference~\cite{Akhmedov:2012ah} to find the average $\nu_\mu$-to-$\mu$ and
$\bar{\nu}_\mu$-to-${\bar \mu}$ cross sections. Modeling the Galactic Center as
an isotropic circle with radius of $5^\circ$ and ignoring detector
angular resolution, we obtain a $5^\circ$ optimal cut, whereas the
solar muons have a $2^\circ$ optimal cut. To find the through-going
muons, we propagate the muons using the parameters of Table (4) of reference~\cite{Chirkin:2004hz}.

In order to generate the energy spectra of the SM particles that are produced at the annihilation point,
we first create the model file for CalcHEP using FeynRules 2.0~\cite{Alloul:2013bka} and then use
CalcHEP 3.6.23~\cite{Belyaev:2012qa} and Pythia-6~\cite{Sjostrand:2006za} to get the spectra.
These spectra are valid both at the Galactic Center and inside the Sun.
In the following we present our result for these two distinct DM annihilation points.
%%%%%%%%%%%%%%%%%%%%%%%%%%%%%%%%%%%%%%%%%%%%%%%%%%%%%%%%%%%%%%%%%%%%
\subsection{Signals from DM annihilation in the Galactic Center}
To calculate the fluxes near the Galactic Center, we run the indirect detection
module of micrOMEGAsv4.1.8~\cite{Belanger:2014vza}
and measure the fluxes for angle of sight $0^{\circ} \leq \Psi \leq 5^{\circ}$.
We use the Navarro-Frenk-White (NFW)~\cite{Navarro:1995iw} profile
$\rho(r)=\rho_0 (r/r_s)^{-1}/(1+r/r_s)^2$ with $r_s=20$ kpc, $\rho_0=0.4$GeV/cm$^3$.
It is noteworthy that if the annihilation
of our proposed DM candidate happens near the Galactic Center, then the final state neutrinos
will encounter oscillations on their way to the Earth.
In the Appendix, we show the flavor dependent probability of neutrino flux that
originates at the Galactic Center and reaches the detector surface on
the Earth. Hence, the final neutrino fluxes will be given by 
Eq.(\ref{final_fluxes}-\ref{osc_prob}). We have checked that varying all neutrino oscillation parameters 
within the 3$\sigma$ allowed range about their best fit central value results in ${\cal O} (3-4\%)$ change 
in the neutrino flux from DM annihilation at the Galactic Center. 
This is much smaller than the difference due to different mass hierarchies.

Following are the interesting features that are seen in Figs. \ref{vd:1ev}, \ref{vd:1evphoton} and \ref{vd:1gev}.
\begin{itemize}
	\item Fig.~\ref{vd:1ev} displays the $\nu_\mu$ fluxes for NH and IH
	for BP1 and BP2 which is for $v_\Delta = 1 ~\rm eV$. As previously argued, at such low vev,
	the triplet scalars generate leptonic final state with almost $100\%$ branching
	ratio and the DM annihilation to two on-shell triplet scalars mainly contribute
	to the annihilation cross section, hence to the $\nu_\mu$ flux. Now, it is to
	be observed that for $m_{\rm DM}$ = 500 GeV, the
	flux rises at around 100 GeV and then falls near 400 GeV and similarly,
	the rise and fall occur at 60 GeV and 640 GeV for $m_{\rm DM}$ = 700 GeV.
	This is not surprising and can be understood by the kinematics of the DM
	annihilation. The triplets are produced on-shell with some boost that
	comes from the mass difference between the DM and the triplets.
	Thus the two-body final states render a box like feature with end points $m_{\rm DM} (1\pm\beta)/2$ where
	$\beta=\sqrt{1-m_{\Delta}^2/m_{\rm DM}^2}$.
	\item We also see from Fig.~\ref{vd:1ev} that there exists a significant difference between
	the NH and the IH scenarios.
	This is due to the fact that in the NH case, we get more taus ($\tau$) from the triplet decays which
	produces $\nu_{\tau}$ and the $\nu_{\tau}$ further gets converted into $\nu_{\mu}$.
	For both the DM mass, the neutrino flux can be easily a factor of two higher in the NH scenario than the IH case.
	The neutrino flux due to degenerate case will be in between the flux generated for NH and IH cases.
	This is simply because in the degenerate case, we get less $\tau$s from the triplet decays
	compared to the NH case but more than the IH case. This is true for all our discussions to follow.
	Therefore, the degenerate case will not provide any new insight and so
    we restrain ourselves from doing any further remark on this case.
 \begin{figure}[ht!]
 	\centering
 	\begin{tabular}{c c}
 	\includegraphics[height=6cm,width=8cm]{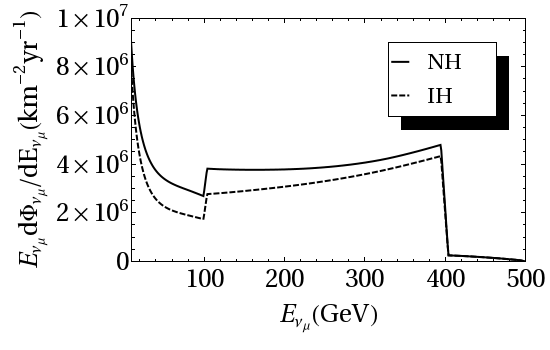} &
 	\includegraphics[height=6cm,width=8cm]{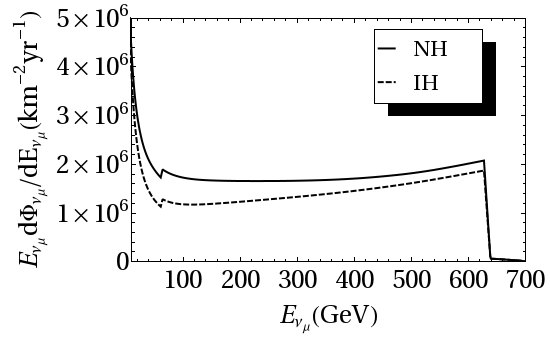}
 	\end{tabular}
 	\caption{Spectra of $\nu_\mu$ from DM annihilation at the Galactic Center
 		 for BP1 (left) with $m_{\rm DM} = 500~\rm GeV$  and BP2 (right) with $m_{\rm DM} = 700 ~\rm GeV$.}
 	\label{vd:1ev}
 \end{figure}
	\item In Fig.~\ref{vd:1evphoton}, we present the diffused photon fluxes for NH and IH cases. In our model,
		the photon flux arises from external charged lepton legs, final state
		radiation, and secondary decay of charged leptons that are directly produced from triplet decay.
	For both the benchmark points, we observe that the photon flux does not distinguish between NH and IH cases~\footnote
%      because the differences are not significant due to the leptonic nature of final states 
{The DM annihilation can also produce photon lines
		via charged scalar loop processes ${\rm DM} + {\rm DM} \to \gamma \gamma$ and ${\rm DM} + {\rm DM} \to Z \gamma$ with respective energies $m_{\rm DM}$ and $m_{\rm DM} (1-\frac{m_Z^2}{m_{\rm DM}^2})$. However, the line signal is highly suppressed relative to the diffused photon signal in our model, which is in agreement with the fact that that Fermi-LAT has not observed any photon line at such energies.}. 
 \begin{figure}[ht!]
 	\centering
 	\begin{tabular}{c c}
 	\includegraphics[height=6cm,width=8cm]{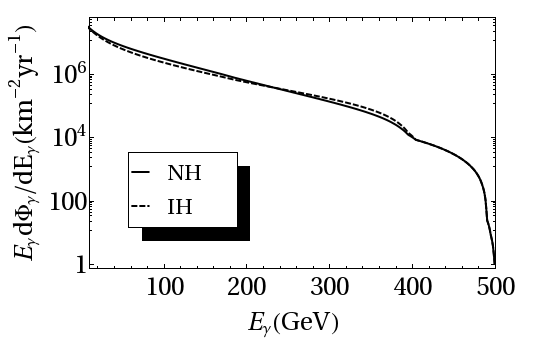} &
 	\includegraphics[height=6cm,width=8cm]{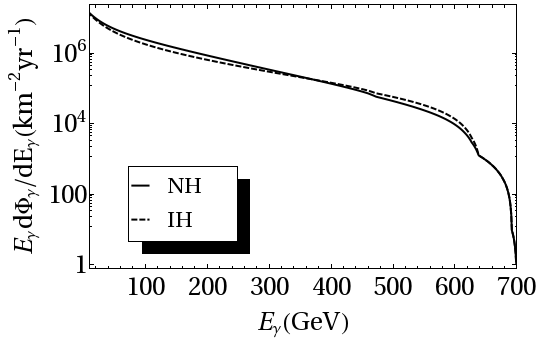}
 	\end{tabular}
 	\caption{Photon spectra from DM annihilation at the Galactic Center for BP1 (left) with $m_{\rm DM} = 500~\rm GeV$ and BP2 (right)
 		with $m_{\rm DM} = 700 ~\rm GeV$.}
 	\label{vd:1evphoton}
 \end{figure}
	\item With the increase in $v_\Delta$, the branching ratio of the triplet
		scalars to leptons reduces and the decay modes with gauge boson final states open up
		and at sufficiently large $v_\Delta$, ${\rm BR}(H^{\pm\pm} \to W^\pm W^\pm) \simeq 100\%$,
		dominating over the leptonic final states.
		Hence, for $v_\Delta=1~\rm GeV$, the dependence on neutrino mass hierarchy gets
		washed away. Moreover, the contribution to neutrino fluxes
		coming from DM annihilating to Higgs will become more pronounced
		in this case. In Fig.~\ref{vd:1gev}, we show neutrino fluxes for BP3 and BP4 and
		evidently these do not distinguish between NH and IH cases. However, the shapes of
		neutrino spectra are different compared to the previous $v_\Delta = 1 ~\rm eV$ cases as shown
		in the Fig.~\ref{vd:1ev}. As expected, the flux is higher in the previous case 
		because the leptonic final states originate from direct triplet decay. The photon flux, however,
		does not draw any discrimination be it between $v_\Delta$=1 eV and 1 GeV cases or between NH and IH scenarios.
	\begin{figure}[ht!]
		\centering
		\begin{tabular}{c c}
		\includegraphics[height=6cm,width=8cm]{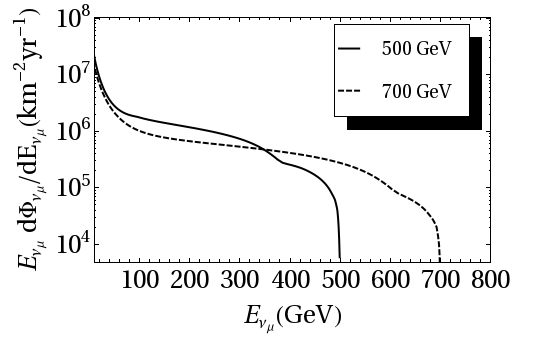} &
		\includegraphics[height=6cm,width=8cm]{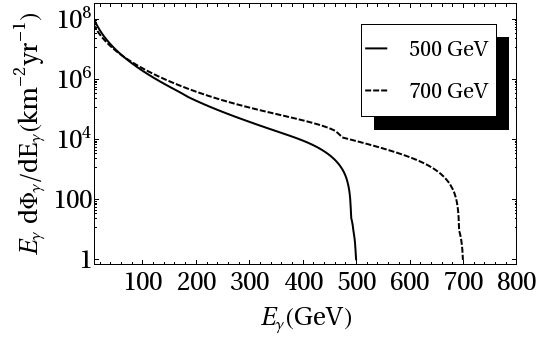}
		\end{tabular}
		\caption{Spectra of $\nu_\mu$ (left) and Photon (right) from DM annihilation at the Galactic Center for BP3 and BP4.}
		\label{vd:1gev}
	\end{figure}
\end{itemize}
%%%%%%%%%%%%%%%%%%%%%%%%%%%%%%%%%%%%%%%%%%%%%%%%%%%%%%%%%
\subsection{Signals from DM annihilations inside the Sun}
DM particles that pass through the Sun lose energy due to the DM-nucleon scattering and get
gravitationally trapped~\cite{Zeldovich:1980st,Silk:1985ax,Krauss:1985ks,Krauss:1985aaa,Press:1985ug,Gaisser:1986ha,Barger:2011em}.
The total rate of DM annihilations $\Gamma_{\rm ann}$ inside the Sun follows~\cite{Jungman:1995df}
\begin{eqnarray} \label{eqAnnihilation}
\Gamma_{\mathrm{ann}} = {C \over 2} {\rm tanh}^2(\sqrt{C A} t) \, .
\end{eqnarray}
Here $C$ is the capture rate of DM particles by the Sun, which is related to $\sigma_{\rm SI}$ in our model,
and $A$ is related to the DM annihilation cross section $\sigma_{\rm ann}$ (for details, see~\cite{Jungman:1995df}).
DM capture and annihilation reach equilibrium when $t > \tau_{\rm eq} \equiv (\sqrt{CA})^{-1}$, in which case $\Gamma_{\rm ann} \approx C/2$.
 For values of $\sigma_{\rm SI}$ given in Table 1, and $\langle \sigma_{\rm ann} v \rangle = 3 \times  10^{-26}$ cm$^3$ s$^{-1}$,
the equilibrium condition is achieved for the age of the Sun. We note that as long as equilibrium is established, the total rate of DM annihilations is controlled by the direct detection cross section $\sigma_{\rm SI}$. Therefore, increasing $\sigma_{\rm ann}$ (which is allowed in scenarios of non-thermal DM) and keeping $\sigma_{\rm SI}$ fixed does not enhance the signal from DM annihilation inside the Sun. It only leads to a faster establishment of equilibrium between DM capture and annihilation.
%It is to be mentioned here that
%the t-channel SM Higgs mediated process will produce the direct detection cross section $(\sigma_{SI})$
%and thus we choose the DM coupling to the Higgs $(\lambda_\Phi)$ at values where
%the cross section stays well below the present LUX~\cite{Akerib:2013tjd} limit.

DM annihilation final states that are most relevant for producing neutrinos inside the
Sun are prompt neutrinos from $H^0/A^0$ and $H^{+}/H^{-}$ decays, $W,Z $ and $\tau$ particles from $H^{+}/H^{-}$ and
$H^{++}/H^{--}$ decays. Other final states (like $e$, $\mu$, and lighter quarks) lose their energy
and stop almost immediately, which results in production of neutrinos at energies below detection threshold.
The charged current interactions inside the Sun convert $\nu_e,~\nu_\mu,~\nu_\tau$ to $e,~\mu,~\tau$ respectively.
Using the charged current neutrino-nucleon cross section~\cite{Akhmedov:2012ah}, we find that neutrino absorption
becomes important (i.e., $L_{\rm abs} < R_{\rm C}$, with $R_{\rm C}$ being the core radius of the Sun)
at energies $E_\nu > 300$ GeV.
For $\nu_e$, the flavor and mass eigenstates are the same deep inside the Sun, and hence
absorption suppresses the flux of $\nu_e$ at energies above $300$ GeV. On the
other hand, $\nu_\tau$ absorption produces $\tau$ particles that decay quickly before 
losing too much energy because of very short lifetime of $\tau$.
This decay produces $\nu_\tau$ at a lower energy, and hence this ``regeneration'' effect populates the spectrum at
energies well below the DM mass.
Since $\nu_\mu$ and $\nu_\tau$ undergo oscillations inside the Sun
that is dominantly set by the atmospheric mass splitting $\Delta m^2_{\rm atm}$, we have $L_{\rm osc} \propto E_\nu/\Delta m^2_{\rm atm}$. 
As long as $L_{\rm abs} \gs L_{\rm osc}/4$, oscillations mix $\nu_\mu$ and $\nu_\tau$ 
efficiently and $\nu_\mu$ final states also feel the regeneration effect. When $L_{\rm abs}$ 
drops below $L_{\rm osc}/4$, which happens at $E_\nu \sim 500$ GeV, oscillations cease to be 
effective and $\nu_\mu$ gets absorbed similar to $\nu_e$. As a consequence, only the $\nu_\tau$ 
final state retains a significant regeneration signature at energies above 
500 GeV.\footnote{Neutrinos also have neutral current interactions with matter inside the Sun that
results in energy loss of the neutrinos from all flavors further shifting their spectra toward lower energies.
However, the cross section for neutral current interactions is a factor of 3 smaller than that for charged current
interactions, which makes them subdominant.}

For neutrinos of sufficiently low energies, vacuum flavor oscillations between the Sun and the Earth are averaged over half a year
when the Sun is below the horizon at the south pole. In particular, below 100 GeV, the oscillation length set by the solar mass
splitting $\Delta m^2_{\rm sol}$ is less than approximately 3 million kilometer change in the Earth-Sun distance over half a year.
The situation changes at energies above 100 GeV where solar neutrino oscillations are not averaged out anymore. 
For monochromatic neutrinos of a single flavor, this significantly affects oscillations between 
$\nu_e$ and $\nu_\mu/\nu_\tau$, and hence the $\nu_\mu$ spectrum at the detector, at high energies.
However, this is not an important effect in our model since DM annihilation produces different neutrino flavors with a continuous energy spectrum in this case. 

We use DarkSUSY 5.1.1~\cite{Gondolo:2004sc} to simulate production of neutrinos in the Sun, their propagation to the South Pole,
and the interaction of $\nu_{\mu}$ with ice at many different energies for all the SM channels.
DarkSUSY does this by interpolating an older WimpSim~\cite{Blennow:2007tw} simulation.
Here we 
%We need to 
run WimpSim 3.05 directly for the prompt neutrino channels 
%since DarkSUSY's interpolation to different energies does
%not work due to the fact that flavor oscillations, as mentioned above, are not averaged out in this case. 
%When running WimpSim
, sample energies spaced at 10 GeV, and use oscillation 
parameters found in~\cite{Forero:2014bxa}. 

We then combine the energy spectra of all the channels at the production 
point with how they propagate, giving us the final neutrino spectra at the 
detector.\footnote{Since ${\bar \nu}$-nucleon cross section is about two 
times smaller than the $\nu$-nucleon cross section
~\cite{Gandhi:1995tf,CooperSarkar:2007cv}, the number of absorbed 
${\bar \nu}$ inside the Sun is smaller. This results in a larger 
${\bar \nu}$ flux at the detector from DM annihilation inside the Sun.}
It should be noted that 
variation of all neutrino oscillation parameters within the $3 \sigma$ allowed range results in ${\cal O} (3-4\%)$ change 
in the final neutrino spectra from the Sun, similar to the Galactic Center case, which is insignificant. 

\begin{figure}[ht!]
	\centering
	\begin{tabular}{c c}
	\includegraphics[height=6cm,width=8cm]{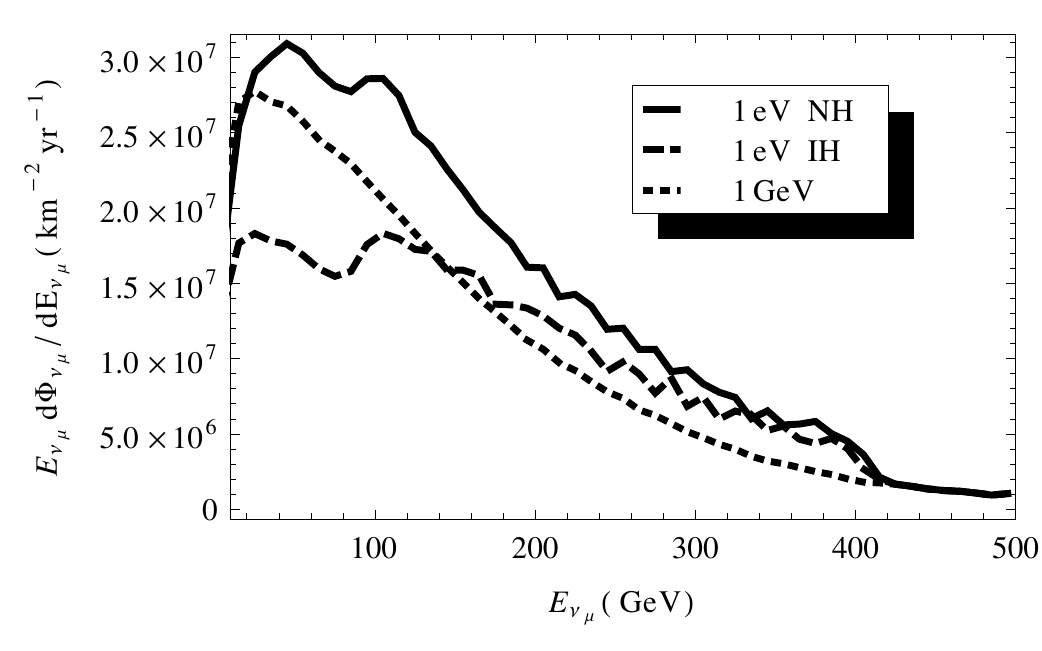} &
	\includegraphics[height=6cm,width=8cm]{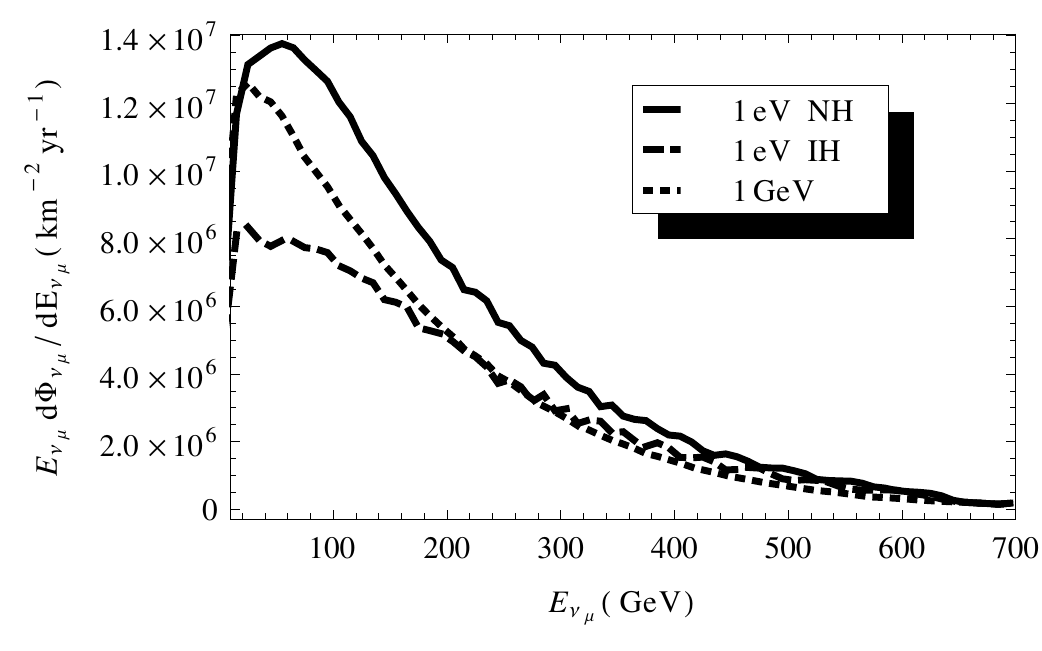}
	\end{tabular}
	\caption{Spectra of $\nu_\mu$ from DM annihilation inside the Sun at the detector for 
		$m_{\rm DM} = 500 ~\rm GeV ~ (700~ \rm GeV)$ in the left (right) panel.}
	\label{nh_absfl_ev}
\end{figure}

In Fig.~\ref{nh_absfl_ev} we show the spectra of $\nu_\mu$ at the detector for $v_\Delta = 1$ eV (BP1, BP2)
and for $v_\Delta = 1$ GeV (BP3) cases. It is seen that at energies below 400 GeV the
NH scenario results in a larger number of $\nu_\mu$ and eventually muons than the IH scenario.
The reason is similar to the Galactic Center case and can be explained by the triplet scalar decay patterns.
For $v_{\Delta} = 1$ eV, as mentioned above, triplet decays mainly produce $\nu_\tau,~\nu_\mu$ and
$\tau,~\mu$ final states in the NH scenario. In the IH scenario, on the other hand,
triplet decays mainly produce $\nu_e$ and $e$ final states. As far
as charged lepton final states are concerned, only $\tau$ decays before losing
a significant fraction of energy and produces neutrinos.
Regarding triplet decays to neutrinos, $\nu_\tau$ (and to a lesser extent $\nu_\mu$)
is the most relevant flavor that survive at energies above $300$
GeV due to regeneration effect. In consequence, for $v_{\Delta} =1$ eV,
a larger number of $\nu_\mu$ arrive at the detector in the NH scenario.
However, the neutrino spectra in the two scenarios do not differ when
$v_{\Delta} = 1$ GeV as in earlier case.
Comparing the neutrino signal from DM annihilation in the Sun (Fig.~\ref{nh_absfl_ev}) and
at the Galactic Center (Fig.~\ref{vd:1ev}), we notice some differences. First, the
kinematic cuts in Fig.~\ref{vd:1ev} are not prominent in Fig.~\ref{nh_absfl_ev}.
This is due to the fact that absorption/regeneration and down scattering inside
the Sun erases sharp features in the spectra. More precisely, only a little bump
develops at low energies for $m_{\rm DM}= 500~\rm GeV$ due to the boosted triplet scalars, 
shown in right panel of Fig.~\ref{nh_absfl_ev},
unlike the box-like structure that can clearly be observed in the Galactic Center neutrino flux. 
At higher energies, instead of a bump only the wiggles can be noticed which are the effect of neutrino
oscillation. Furthermore, the difference between
the spectra in NH and IH scenarios in Fig.~\ref{nh_absfl_ev} is smaller than that
in Fig.~\ref{vd:1ev}. This is because stopping of muons and partial absoprtion of $\nu_\mu$ inside the Sun 
decreases the overall flux of $\nu_\mu$ arriving at the detector.
Nevertheless, the NH scenario yields a larger flux than the IH scenario when $v_{\Delta} = 1$ eV.

We emphasize that the neutrino signal from DM annihilation inside the Sun is
complementary to that from the Galactic DM annihilation as they are set by different cross sections 
($\sigma_{\rm SI}$ and $\sigma_{\rm ann}$ respectively).
%Comparing 
We see from Figs.~\ref{vd:1ev} and \ref{nh_absfl_ev} that for the model parameters given in Table 1
%we see that 
the signal from the Sun is about an order of magnitude
stronger than that from the Galactic Center. As we will discuss later, both of these signals can be enhanced further.
%if larger values of $\lambda_\phi$ and $\lambda_\Delta$ are chosen, see Table 1, which we will disucss later.}   
%even though $\sigma_{\rm SI}$ is well below the current LUX bound.
%The ratio will be even larger if the annihilation rate is smaller
%than the thermal freeze-out value.
%
\subsection{Muon spectra at the detector}
\label{muon_spec}
We now present the muon spectra obtained from the conversion of $\nu_\mu$ at the detector, which is the observed signal at neutrino telescopes.
For simulation purpose, we assume a detector
that has the same capability as the IceCube DeepCore array.
	\begin{figure}[ht!]
		\centering
		\begin{tabular}{c c}
		\includegraphics[height=6cm,width=8cm]{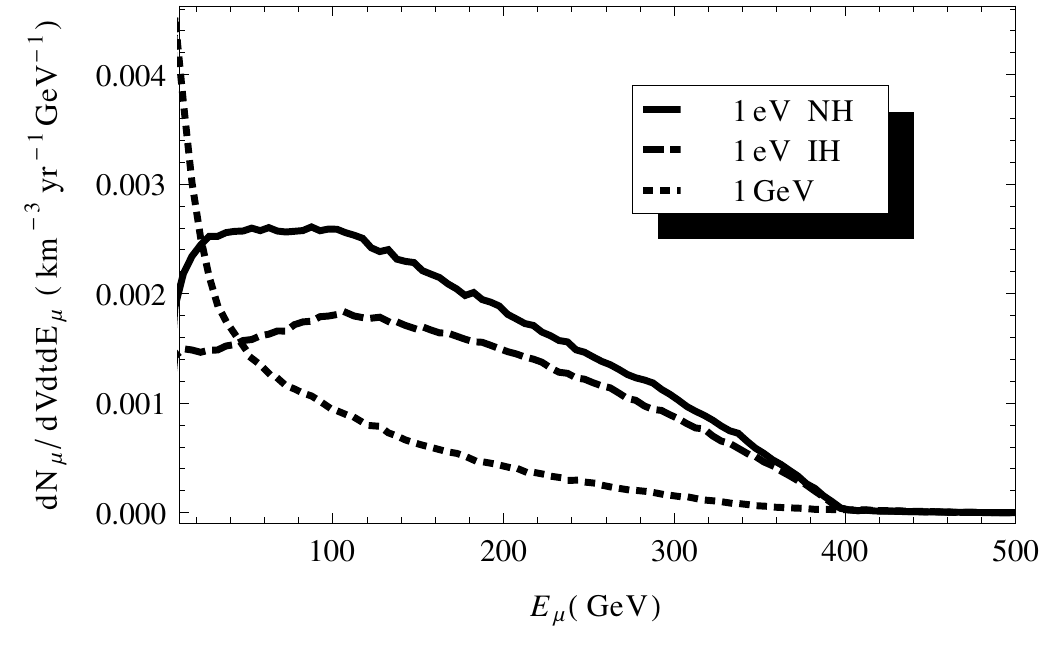} &
		\includegraphics[height=6cm,width=8cm]{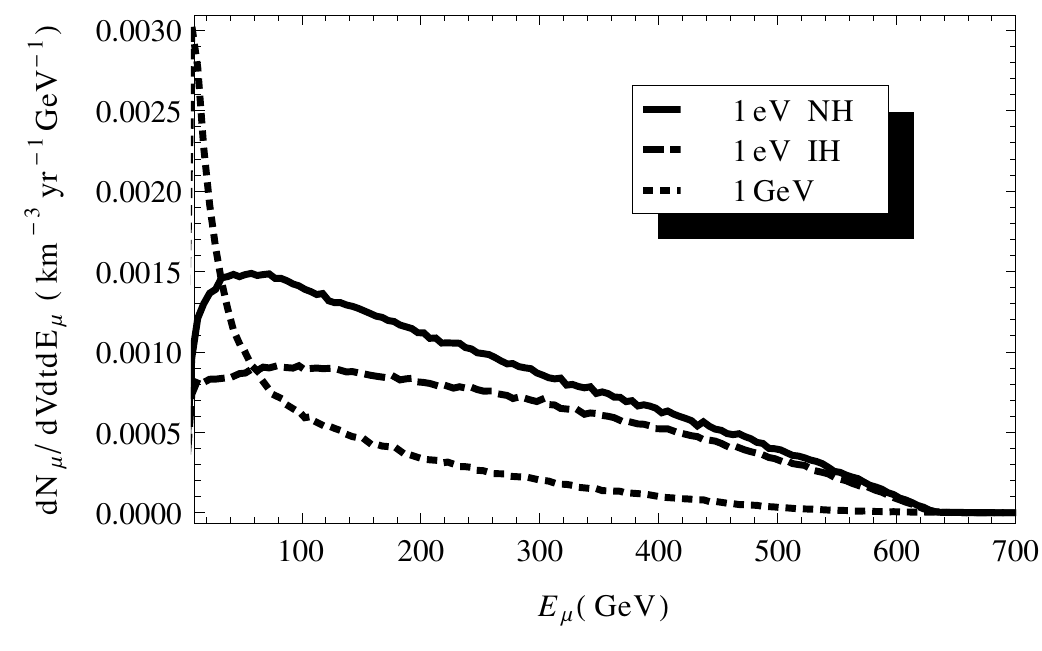}\\
		\includegraphics[height=6cm,width=8cm]{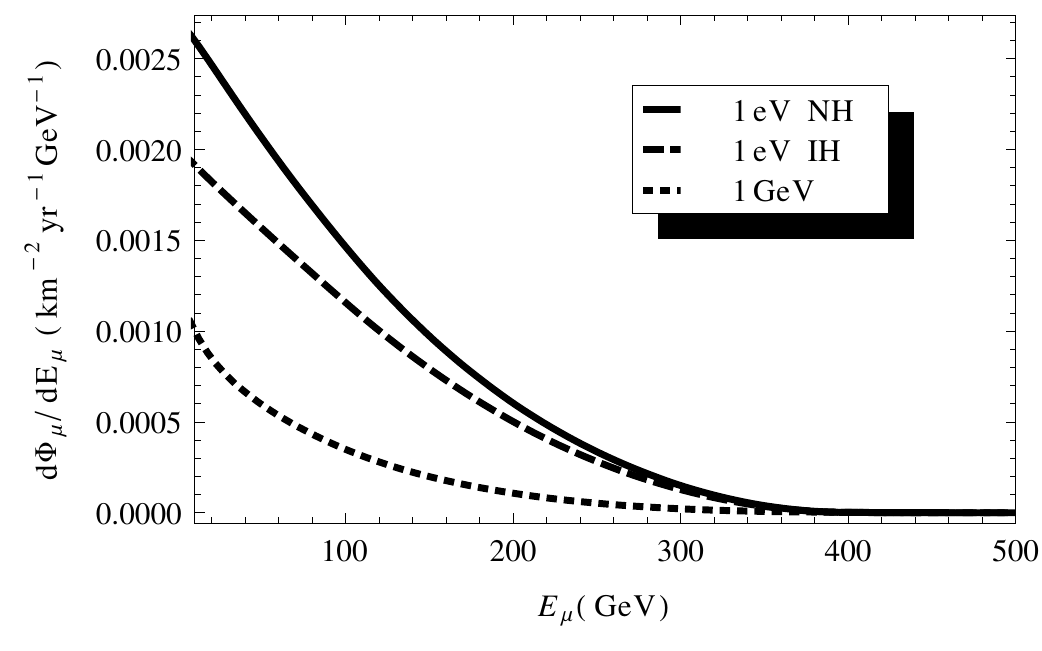} &
		\includegraphics[height=6cm,width=8cm]{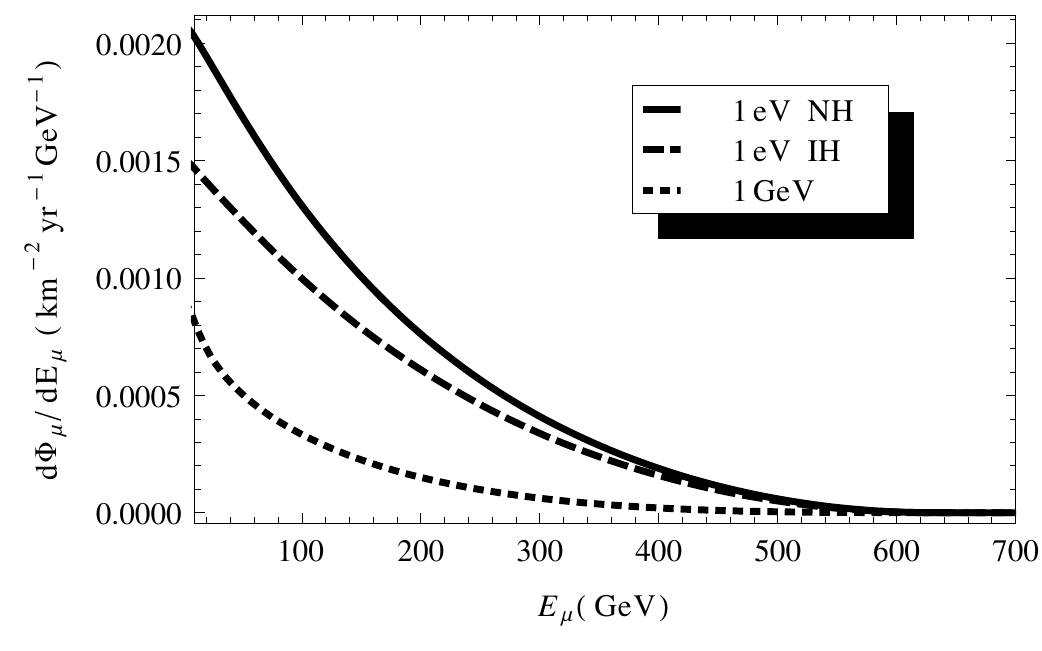}
	\end{tabular}
		\caption{Spectra of muons from DM annihilation at the Galactic Center
			for $m_{\rm DM} = 500 ~\rm GeV ~ (700~ \rm GeV)$ in the left (right) panel. 
			Upper and lower panels show contained and through-going muon spectra with an angular cut of $5^{\circ}$ respectively.}
		\label{galacticcenter}
	\end{figure}
In Fig.~\ref{galacticcenter},
	we show the spectra of contained and through-going muons in the detector
	from DM annihilation at the Galactic Center. In this case,
	the signal comes from a region around the Galactic Center that has $5^{\circ}$
	angular extension. Hence, in order to optimize the signal to background ratio,
	we have imposed a $5^{\circ}$ angular cut on the muons relative to
	the center of the galaxy.
 
\begin{figure}[ht!]
	\centering
	\begin{tabular}{c c}
		\includegraphics[height=6cm,width=8cm]{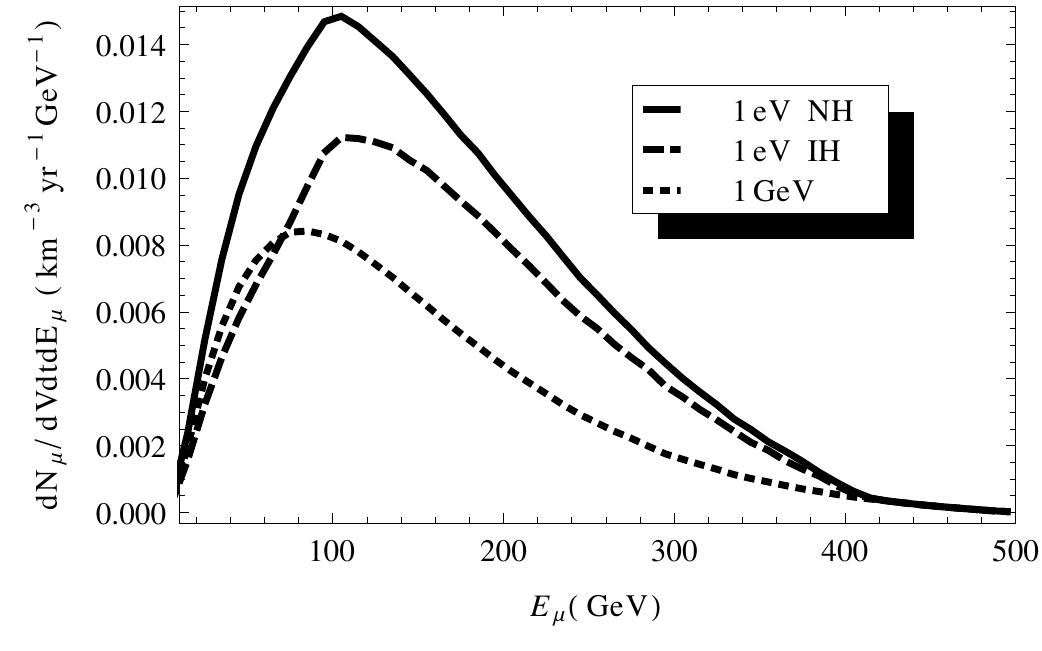} &
		\includegraphics[height=6cm,width=8cm]{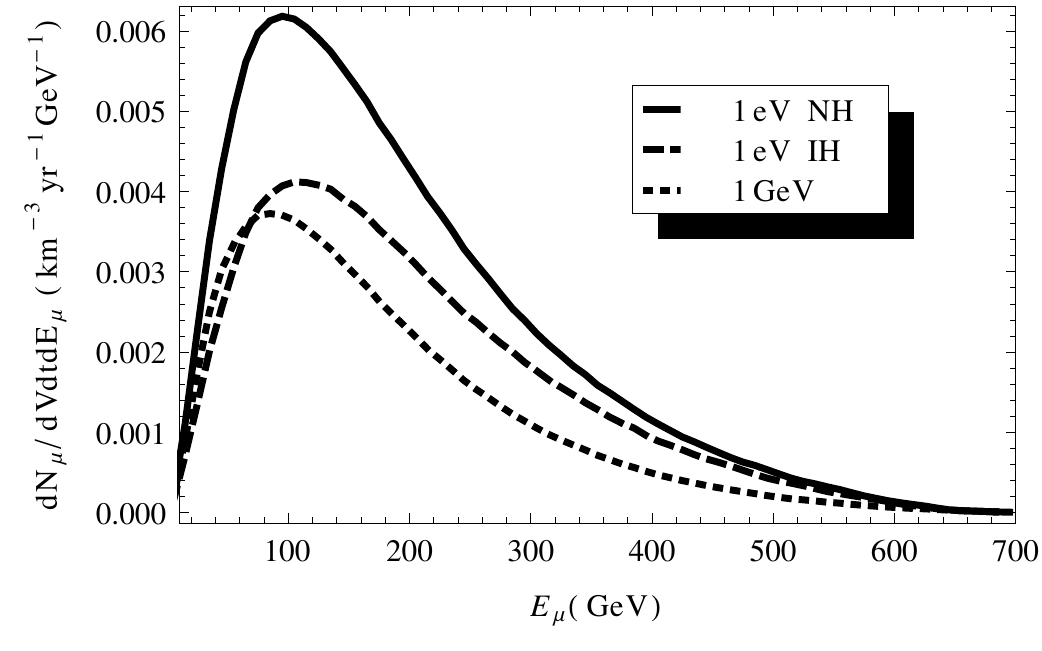}\\
		\includegraphics[height=6cm,width=8cm]{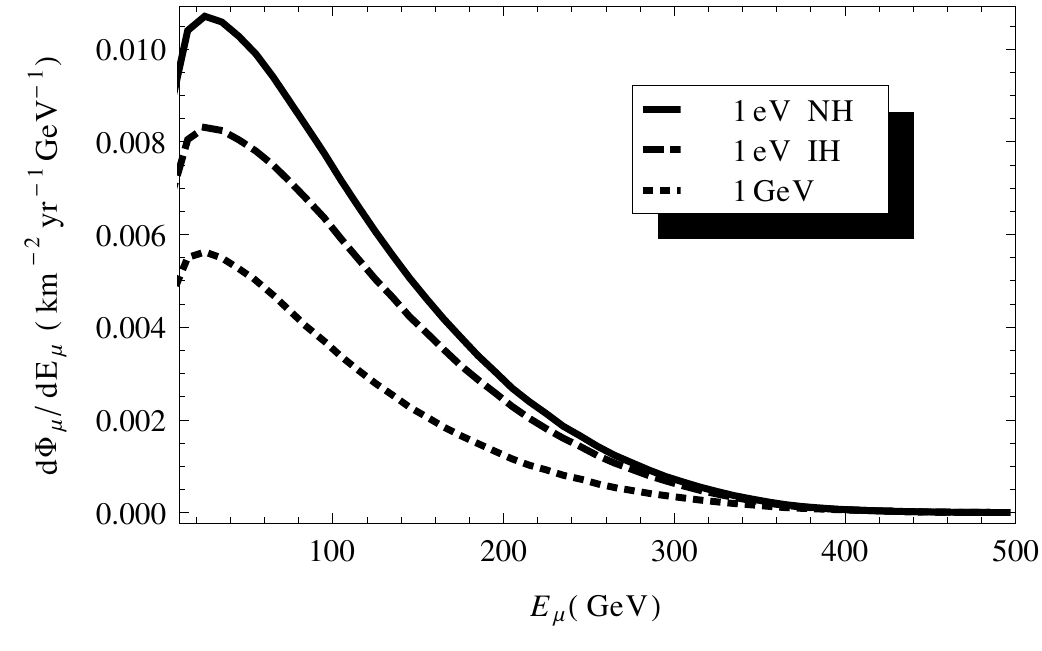} &
		\includegraphics[height=6cm,width=8cm]{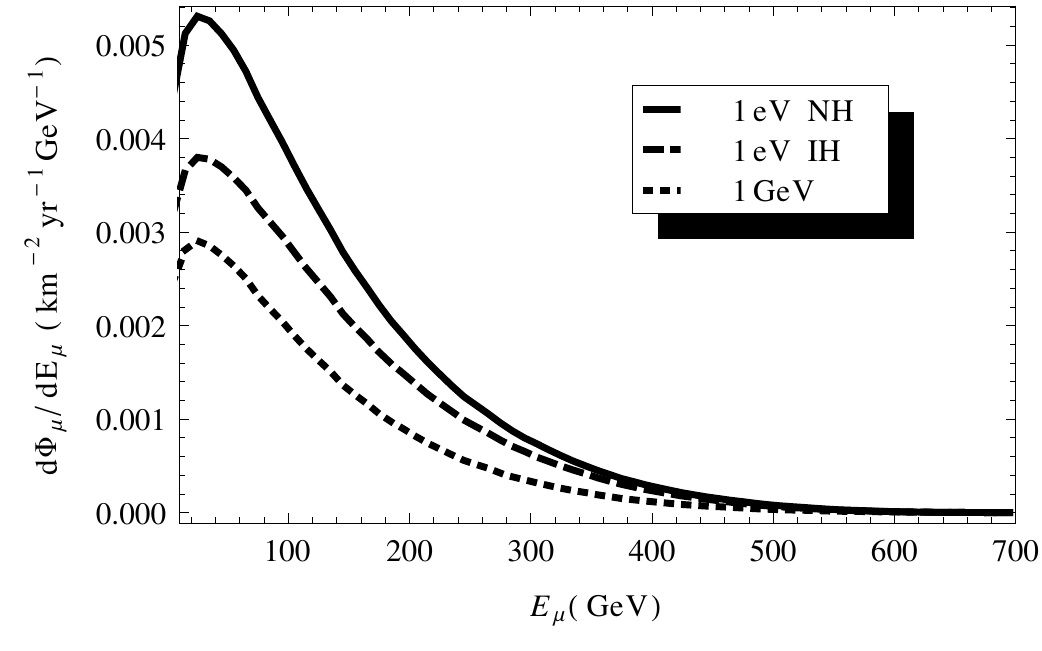}
	\end{tabular}
	\caption{Spectra of muons from DM annihilation in the Sun for 
		$m_{\rm DM} = 500 ~\rm GeV ~ (700~ \rm GeV)$ in the left(right) panel. 
		Upper and lower panels show contained and through-going muon spectra with an angular cut of $2^{\circ}$ respectively.}
	\label{sun}
\end{figure}

In Fig.~\ref{sun}, we show the
	spectra of contained and through-going muons from DM annihilation inside
	the Sun. Since the Sun is a point-like source,
	the optimal signal to background ratio is obtained for an angular cut of $2^{\circ}$, 
	in this case. The peaks in both figures are due to the imposed angular cuts 
	that eliminate muons produced from $\nu_\mu$ conversion below a certain energy. 
	For through-going muons the peak occurs at a lower energy, which can be 
	understood by noting that the measured energy of a through-going muon is 
	in general less than the actual energy at the production point.   

As expected, the muon spectra have less features than the neutrino spectra shown in Figs.~\ref{vd:1ev}-\ref{nh_absfl_ev}.
Nevertheless, the difference between $v_{\Delta} = 1$ eV and $v_{\Delta}= 1$ GeV cases, as well as the NH and IH
scenarios in the former case, are clearly visible in the muon spectra arising from both
the Galactic Center and the Sun. Similar to the neutrino spectra, substantial deviation among the cases is seen
in the signal arising from DM annihilation at the Galactic Center than in the Sun.
However, we also note that the absolute value of the flux from the Galactic Center is smaller by
an order of magnitude than that from the Sun.

The muon spectra shown in Figs.~\ref{galacticcenter} and \ref{sun} are obtained assuming the nominal thermal 
freeze-out value of $\left<\sigma_{\rm ann} v \right>$. However, the Fermi-LAT data from Milky Way dwarf spheroidal 
galaxies~\cite{Ackermann:2015zua} allows larger values on the DM annihilation cross 
section for the values of DM mass that we have considered here. One can see from Fig. 8 of~\cite{Ackermann:2015zua} that for the $b {\bar b}$ and $W^{\pm} W^{\mp}$ final states $\langle \sigma_{\rm ann} v \rangle$ can be larger than the thermal freeze-out value by a factor of 4 (6) when $m_{\rm DM} = 500 ~ (700)$ GeV. In BP1 and BP2, where the distinction between the NH and IH scenarios is significant, these final states arise from DM annihilation to the SM Higgs that is controlled by the coupling $\lambda_\Phi$ of Eq.~(\ref{dm_potential}). As explained before, $\lambda_\Phi$ also controls $\sigma_{\rm SI}$ in our model. Hence raising $\langle \sigma_{\rm ann} v \rangle$ by the above factors, due to a larger value of $\lambda_\Phi$, also results in an increase in $\sigma_{\rm SI}$ compared with that given in Table 1, which is still consistent with the LUX bounds~\cite{Akerib:2013tjd}. For the $\tau^+ \tau^-$ final state, $\langle \sigma_{\rm ann} v \rangle$ can be larger than the thermal freeze-out value by a factor of 10 (17) when $m_{\rm DM} = 500 ~ (700)$ GeV~\footnote{  
%(10), for $m_{\rm DM}= 500$ GeV, and by a factor of 6 (17), for $m_{\rm DM}= 700$ GeV, for 
%$b \bar b ~ (\tau^+ \tau^-)$ final states respectively~\footnote{
It should be noted that the Fermi-LAT bound cannot be directly applied to our model since 
%on the annihilation cross section given by
%the Fermi-LAT experiment can not be directly imposed on our scenario. 
%As an explanation, it should be noted that the DM candidate of our model 
we have four-body annihilation final states in this case.
%annihilates into four body final states instead of two-body final states as assumed by the Fermi-LAT in their analysis. 
Moreover, we have a combination of different final states instead of 100$\%$ of just one final state. 
%Secondly, in our analysis, the branching ratio 
%of DM annihilation into each four body final states is model dependent and much less than 
%100\% unlike the analysis of Fermi-LAT where they assume the DM annihilates into individual final states with 100\% branching ratio.
Nevertheless, the Fermi-LAT limit provides a reasonable upper bound on $\sigma_{\rm ann}$ in this case too.}. This final state arises due to DM annihilation to the triplet Higgs,
which is controlled by the coupling $\lambda_\Delta$ of Eq.~(\ref{dm_potential}).  
Increasing $\lambda_{\Delta}$ results in an increase in $\sigma_{\rm ann}$, but does not affect $\sigma_{\rm SI}$.

%For non-leptonic channels, specifically $W^\pm W^\mp$ or Higgs final 
%states (the latter mainly decaying to $b {\bar b}$), the bound on $\sigma_{\rm ann}$ 
%is a factor of few larger than the thermal freeze-out value for DM masses that we have considered here.
%In BP1 and BP2, where the distinction between the NH and IH scenarios is significant, 
%the above mentioned final states only arise from DM annihilation to SM Higgs controlled by the coupling 
%$\lambda_\Phi$ of Eq.~(\ref{dm_potential}). As explained before, $\lambda_\Phi$ also controls 
%$\sigma_{\rm SI}$ in our model, hence, increasing $\lambda_\Phi$ results in an 
%increase in both $\sigma_{\rm ann}$ and $\sigma_{\rm SI}$.  
%On the other hand, for leptonic final states, the gamma-ray constraint on $\sigma_{\rm ann}$ 
%is larger than the thermal freeze-out value by an order of magnitude or so. 
%These leptonic final states arise due to DM annihilation to the triplet Higgs,
%which is controlled by the coupling $\lambda_\Delta$ of Eq.~(\ref{dm_potential}).  
%Increasing $\lambda_{\Delta}$ results in an increase in $\sigma_{\rm ann}$, but does not affect $\sigma_{\rm SI}$.  

\begin{figure}[htbp!]
		\centering
		\begin{tabular}{c c}
			\includegraphics[height=6cm,width=8cm]{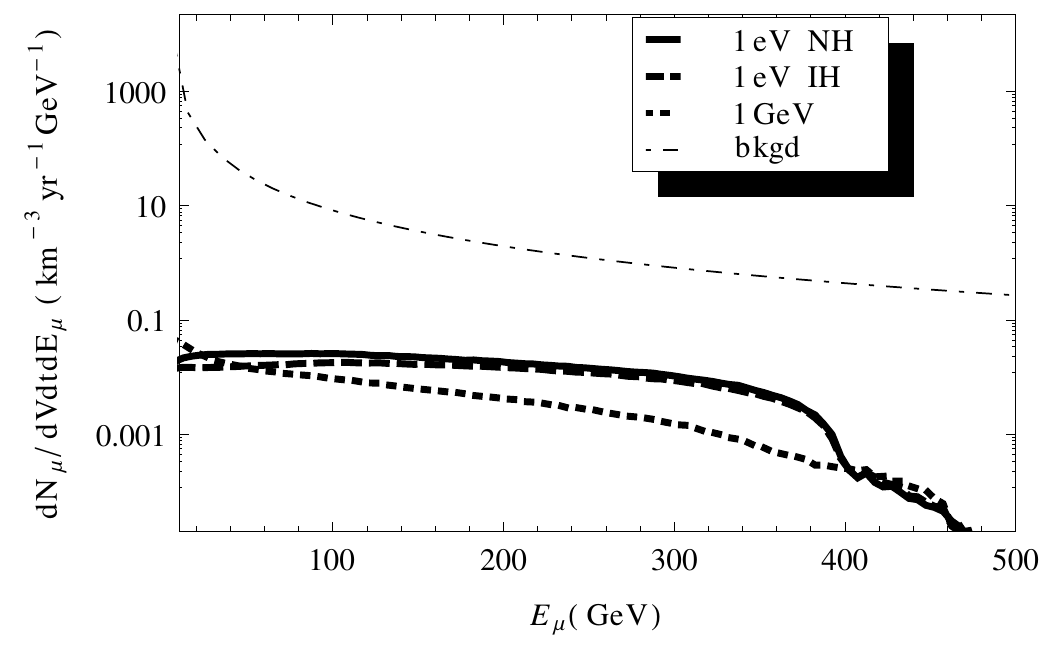} &
			\includegraphics[height=6cm,width=8cm]{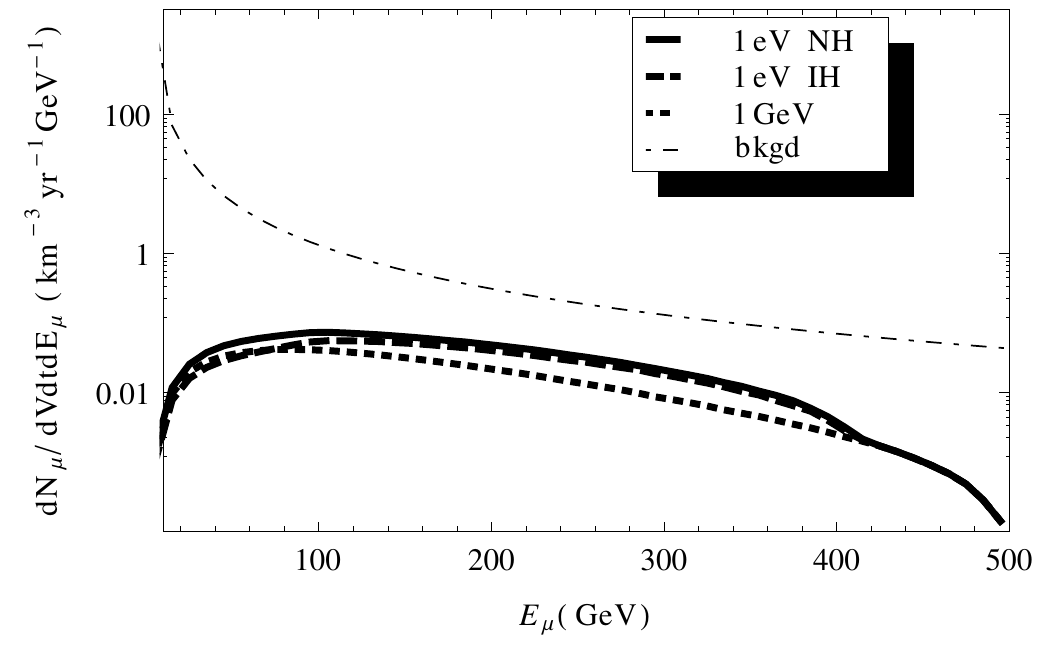}
		\end{tabular}
		\caption{Spectra of contained muons from DM annihilation at the Galactic Center (left) and inside the Sun (right)
			for $m_{\rm DM} = 500 ~\rm GeV$. The value of $\sigma_{\rm ann}$ used to obtain the muon flux is just below the current bounds from Fermi-LAT. The top line in both figures is due to the background arising from the atmospheric neutrinos.}
		\label{bkgdcomparison}
	\end{figure}

Therefore, by invoking non-thermal mechanisms for DM production in 
the early universe, we can obtain considerably larger neutrino fluxes from 
DM annihilation at the Galactic Center and inside the Sun.
%We can raise $\sigma_{\rm ann}$ and $\sigma_{\rm SI}$ both by simltaneously increasing $\lambda_\Delta$ and $\lambda_\Phi$ consistent with the %Fermi-LAT limit~\cite{Ackermann:2015zua} and the LUX bound~\cite{Akerib:2013tjd}. 
%One can see from Fig. 8 of~\cite{Ackermann:2015zua} that $\sigma_{\rm ann}$ can be increased by a factor of 4 (10), for $m_{\rm DM}= 500$ GeV, and %by a factor of 6 (17), for $m_{\rm DM}= 700$ GeV, for 
%$b \bar b ~ (\tau^+ \tau^-)$ final states respectively~\footnote{It should be noted that the Fermi-LAT bound cannot be directly applied to our model since 
%on the annihilation cross section given by
%the Fermi-LAT experiment can not be directly imposed on our scenario. 
%As an explanation, it should be noted that the DM candidate of our model 
%we have four-body annihilation final states in this case.
%annihilates into four body final states instead of two-body final states as assumed by the Fermi-LAT in their analysis. 
%Moreover, we have a combination of different final states instead of 100$\%$ of just one final state. 
%Secondly, in our analysis, the branching ratio 
%of DM annihilation into each four body final states is model dependent and much less than 
%100\% unlike the analysis of Fermi-LAT where they assume the DM annihilates into individual final states with 100\% branching ratio. 
%Nevertheless, the Fermi-LAT limit provides a reasonable upper bound on $\sigma_{\rm ann}$ in this case too.}. 
The resulting enhancement in the muon events, compared to Figs.~\ref{galacticcenter} and \ref{sun}, leads to a better prospect for detection of the neutrino signal against the atmospheric background.
In Fig.~\ref{bkgdcomparison}, we show the spectra of contained muons due to DM annihilation 
at the Galactic Center (left panel) and inside the Sun (right panel) for annihilation cross sections that are just below the Fermi-LAT limits, where $m_{\rm DM} = 500$ GeV and the background from atmospheric neutrinos is also shown for comparison. Contained muons from DM annihilation 
inside the Sun provide the best detection opportunity. 
Within energy interval 100-400 GeV, the total number of contained muons for the NH and IH scenarios is 11 and 9 respectively (compared with 95 for the background). The maximum signal to background ratio, 17$\%$ and 15$\%$ for the NH and IH scenarios respectively, occurs at an energy of 255 GeV.
%, for an upper energy cut of 255 GeV on the muon spectra.  
%255 contained muons from the atmospheric background. 
For a neutrino telescope with the same capability 
as the IceCube DeepCore array, 3$\sigma$ discovery of NH and IH scenarios takes 8 and 12 years respectively. 
To distinguish the different neutrino mass hierarchies, one should go beyond the simple number count 
and perform a careful shape analysis, which is beyond the scope of this paper.

There are proposals for directly measuring the mass hierarchy by using atmospheric neutrinos in future extensions of neutrino telescopes such as PINGU (Precision IceCube Next Generation Upgrade)~\cite{Cowen:2014rga}. It is possible to exclude the wrong mass ordering by this method (as well as in neutrino beam experiments) at the 3$\sigma$ level within the next 10-15 years~\cite{Blennow:2013oma}. Our approach, which exploits an interesting connection between DM and neutrinos, is complementary to this direct method and makes another case for the future neutrino telescopes with much larger volume like IceCube-Gen2~\cite{Aartsen:2014njl}. 
%has been investigated and it is found that the wrong mass ordering will be excluded at 3$\sigma$ C.L. within the next 10 to 15 %years~\cite{Blennow:2013oma}. Since we are studying  neutrino physics from the dark matter decay, our approach is compimentary and combining results %from both approaches will help us to understand both dark matter and neutrinos.       
	
%The left panel shows the comparison of the contained muons
%due to DM annihilation at the Galactic center with the atmospheric background.
%One should note that the neutrino flux from DM annihilation inside the Sun
%is set by the DM-nucleon scattering cross section ($\sigma_{SI}$),
%which depends on the coupling $\lambda_\Phi$. This coupling also controls
%the DM annihilation to the SM Higgs, which 
%is subject to tight constraints set by Fermi as stated earlier.
%Therefore, the muon signal from DM annihilation in the Sun cannot be significantly enhanced,
%which is the reason why there is no band in the right panel of Fig.~\ref{bkgdcomparison}.

%%%%%%%%%%%%%%%%%%%%%%%%%%%%%%%%%%%%%%%%%%%%5
\section{Summary and conclusions}\label{conclusion}
In this paper, we have considered an extension of the SM that explains neutrino masses and mixing angles 
via the type-II seesaw mechanism
and includes a singlet scalar as a viable DM
candidate. 
The DM interacts with the triplet scalar that generates light neutrino masses thereby linking the two sectors.
We have studied the neutrino signal from DM annihilation
at the Galactic Center and inside the Sun in this model that could be detected at IceCube. 
Our main results are summarized as follows:
\begin{itemize}
	\item In the region of the parameter
     space where the triplet scalar dominantly decays to leptonic final states, 
     the flux of $\nu_\mu$ from DM annihilation depends upon the
     neutrino mass hierarchy (normal, inverted or degenerate). 
     The photon flux, on the other hand,
     is practically the same for different hierarchies.
     
     \item The difference in the flux of $\nu_\mu$ at the detector for NH and IH 
     is visible. 
     The NH scenario produces a larger flux because of its
     mass pattern. This holds for both signals from DM annihilation at the Galactic Center and  
     inside the Sun. 
     
     \item The difference is more significant in the neutrino signal arising from DM annihilation at the Galactic Center
     than that arises from DM annihilation inside the Sun.
     This is mainly because the interaction of neutrinos with matter inside the Sun
     results in a moderate attenuation of the final $\nu_\mu$ flux at the detector.  
    
     \item 
     The same reason holds for the contained and through-going muons 
     that are produced from the conversion of $\nu_\mu$s at the detector. 
     The muon spectra have less features than the neutrino spectra. 
     Nevertheless, these muons can be detected over the atmospheric background. 
     %by a detector that has the same capability as the IceCube DeepCore array with multiyear data.  
         
     \item For annihilation cross sections just below the Fermi-LAT limits, 
     future extensions of IceCube are able to discover the neutrino signal 
     and distinguish the different neutrino mass hierarchies with multiyear data. 
%If the constraint arising from the relic density is relaxed by assuming non-thermal DM production, 
%then the $\nu_\mu$ flux from the Galactic Center can be enhanced by an order of magnitude.
\end{itemize}
The LHC in tandem with
direct detection experiments, the IceCube and Fermi-LAT will be able to probe this minimal 
extension of the SM that utilizes a new Higgs sector. 
Finally, we would like to conclude by pointing out the fact that this analysis can be
applied to other models that link DM to the neutrino sector.
%%%%%%%%%%%%%%%%%%%%%%%%%%%%%%%%%%%%%%%%%%5
\section{Acknowledgements}
This work is supported in part by NSF Grant
No. PHY-1417510 (R.A. and B.K.) and DOE Grant DE-FG02-13ER42020 (B.D.). We thank
the International Center for Theoretical Physics (ICTP), Trieste, Italy,
where this work was initiated. D.K.G. and I.S. would like to thank Bhupal Dev for
useful discussions. D.K.G also would like to thank Theory Unit of the Physics
Department, CERN, Geneva, Switzerland where part of this work was done.
R.A. and B.K. would like to thank Shashank Shalgar for valuable discussions.
R.A. and B.D. thank the Center for Theoretical Underground Physics and 
Related Areas (CETUP* 2015) for its hospitality and for partial support 
during the completion of this work.
%%%%%%%%%%%%%%%%%%%%%%%%%%%%%%%%%%%%%%%%%%%%%%%%%%%%%%%%%%%%%%%%%%%%%%
\appendix
\section*{Appendix}
\section{Oscillation Probability Calculation}
\label{oscillation}
The flavor eigenstate of neutrinos can be related to the mass eigenstate
by the $3 \times 3$ Pontecorvo-Maki-Nakagawa-Sakata (PMNS) mixing matrix.
\begin{eqnarray}
\left(\begin{array}{c}
\nu_e \\
\nu_\mu \\
\nu_\tau
\end{array}\right) = \left(\begin{array}{ccc}
U_{e_1} & U_{e_2} & U_{e_3}\\
U_{\mu_1} & U_{\mu_2} & U_{\mu_3}\\
U_{\tau_1} & U_{\tau_2} & U_{\tau_3}
\end{array}\right)\left(\begin{array}{c}
\nu_1 \\
\nu_2 \\
\nu_3
\end{array}\right)
\end{eqnarray}
The neutrinos from the Galactic Center encounter flavor oscillation while
reaching the Earth. However, since it traverses very large distance with respect
to its oscillation length, hence the oscillation term is averaged out. The probability
of one neutrino flavor to oscillate into other becomes simply,
\begin{subequations}
\begin{eqnarray}
<P(\nu_\alpha \to \nu_\beta)> &=& \sum_{i=1}^{3}|U_{\alpha_i}|^2 {|U_{\beta_i}|^2} \\
&=& XX^T
\label{prob:eq} \\
\rm with,~~~&& \nonumber \\
X &=& \left(\begin{array}{ccc}
|U_{e_1}|^2 & |U_{e_2}|^2 & |U_{e_3}|^2\\
|U_{\mu_1}|^2 & |U_{\mu_2}|^2 & |U_{\mu_3}|^2\\
|U_{\tau_1}|^2 & |U_{\tau_2}|^2 & |U_{\tau_3}|^2
\end{array}\right)
\label{xval}
\end{eqnarray}
\end{subequations}
The PMNS matrix is usually parametrized in terms of the three mixing angles
$\theta_{12} ; \theta_{23} ; \theta_{13}$ , and one Dirac $(\delta)$ and two Majorana $(\alpha_1 ; \alpha_2)$
CP phases:
\begin{eqnarray}
U = \left(\begin{array}{ccc}
c_{12}c_{13} & s_{12}c_{13} & s_{13}e^{-i\delta}\\
-s_{12}c_{23}-c_{12}s_{23}s_{13}e^{i\delta} &
c_{12}c_{23}-s_{12}s_{23}s_{13}e^{i\delta} & s_{23}c_{13}\\
s_{12}s_{23}-c_{12}c_{23}s_{13}e^{i\delta} &
-c_{12}s_{23}-s_{12}c_{23}s_{13}e^{i\delta} & c_{23}c_{13}
\end{array}\right)\times{\rm
  diag}(e^{i\alpha_1/2},e^{i\alpha_2/2},1)
\label{pmns:mat}
\end{eqnarray}
The oscillated flux of the neutrinos at the detector at earth,
\begin{eqnarray}
\left(\begin{array}{c}
\Phi_{\nu_e} \\
\Phi_{\nu_\mu} \\
\Phi_{\nu_\tau}
\end{array}\right) = \left(\begin{array}{ccc}
P_{11} & P_{12} & P_{13}\\
P_{21} & P_{22} & P_{23}\\
P_{31} & P_{32} & P_{33}
\end{array}\right)\left(\begin{array}{c}
\Phi_{\nu_e}^0 \\
\Phi_{\nu_\mu}^0 \\
\Phi_{\nu_\tau}^0
\end{array}\right)
\label{final_fluxes}
\end{eqnarray}
From Eq.~(\ref{prob:eq}) and Eq.~(\ref{xval}),
\begin{subequations}
\begin{eqnarray}
P_{11} &=& |U_{e_1}|^4 + |U_{e_2}|^4 + |U_{e_3}|^4  \\
P_{12} &=& |U_{e_1}|^2|U_{\mu_1}|^2 + |U_{e_2}|^2|U_{\mu_2}|^2 + |U_{e_3}|^2|U_{\mu_3}|^2 = P_{21}  \\
P_{13} &=& |U_{e_1}|^2|U_{\tau_1}|^2 + |U_{e_2}|^2|U_{\tau_2}|^2 + |U_{e_3}|^2|U_{\tau_3}|^2 = P_{31} \\
P_{22} &=& |U_{\mu_1}|^4 + |U_{\mu_2}|^4 + |U_{\mu_3}|^4  \\
P_{23} &=& |U_{\mu_1}|^2|U_{\tau_1}|^2 + |U_{\mu_2}|^2|U_{\tau_2}|^2 + |U_{\mu_3}|^2|U_{\tau_3}|^2 = P_{32}  \\
P_{33} &=& |U_{\tau_1}|^4 + |U_{\tau_2}|^4 + |U_{\tau_3}|^4
\end{eqnarray}
\end{subequations}
We use the latest global analysis of the 3-neutrino oscillation data 
\cite{Forero:2014bxa},
 \begin{eqnarray}
\Delta m^2_{\rm 21} =  7.60\times 10^{-5}~{\rm eV}^2\,, & \theta_{12} = 34.6^\circ\,, \nonumber \\ 
\Delta m^2_{\rm 31} = 2.48\times 10^{-3}~{\rm eV}^2 ~ (\rm NH )\,, &
\Delta m^2_{\rm 31} = 2.38\times 10^{-3}~{\rm eV}^2 ~ (\rm IH )\,,~ \nonumber \\
\delta = 1.41\pi ~ (\rm NH )\,, & \delta = 1.48\pi ~ (\rm IH )\,, \nonumber \\
\theta_{23} = 48.9^\circ~(\rm NH ) \,, &  ~\theta_{23} = 49.2^\circ~{(\rm IH)} \,, \nonumber \\
\theta_{13} = 8.6^\circ~ (\rm NH )\, & \theta_{13} = 8.7^\circ ~{(\rm IH)}\,.
 \label{neu_osc_data}
 \end{eqnarray}
%
\begin{comment}
\begin{subequations}
 \begin{eqnarray}
%\theta_{12} &=& 34.4^{\circ}; ~\theta_{23}=40.8^{\circ};~\theta_{13}=9^{\circ}~\rm and \delta = 0.8\pi;
\label{neu_osc_data}
{\rm Normal~ Hierarchy ~:~} P &=& \left(\begin{array}{ccc}
 0.5477 & 0.2764 & 0.1759 \\
0.2764 & 0.3490 & 0.3746 \\
0.1759 & 0.3746 & 0.4495
\end{array}\right) \\
%\theta_{12} &=& 34.4^{\circ}; ~\theta_{23}=40.8^{\circ};~\theta_{13}=9^{\circ}~\rm and \delta = 0.8\pi;
{\rm Inverted~ Hierarchy ~:~} P &=& \left(\begin{array}{ccc}
0.5471 & 0.2702 & 0.1827 \\
0.2702 & 0.3537 & 0.3761 \\
0.1827 & 0.3761 & 0.4412
\end{array}\right)
\label{prob_mat}
 \end{eqnarray}
\end{subequations}
\end{comment}
\begin{subequations}
	\begin{eqnarray}
	%\theta_{12} &=& 34.4^{\circ}; ~\theta_{23}=40.8^{\circ};~\theta_{13}=9^{\circ}~\rm and \delta = 0.8\pi;
	{\rm Normal~ Hierarchy ~:~} P &=& \left(\begin{array}{ccc}
	0.5380 & 0.1979 & 0.2640 \\
	0.1979 & 0.4240 & 0.3779 \\
	0.2640 & 0.3779 & 0.3579
	\end{array}\right) \\
	%\theta_{12} &=& 34.4^{\circ}; ~\theta_{23}=40.8^{\circ};~\theta_{13}=9^{\circ}~\rm and \delta = 0.8\pi;
	{\rm Inverted~ Hierarchy ~:~} P &=& \left(\begin{array}{ccc}
	0.5377 & 0.2009 & 0.2613 \\
	0.2009 & 0.4223 & 0.3766 \\
	0.2613 & 0.3766 & 0.3620
	\end{array}\right)
        \label{osc_prob}
	\end{eqnarray}
\end{subequations}

%%%%%%%%%%%%%%%%%%%%%%%%%%%%%%%%%%%%%%%%%%%%%%%%%%%%%%%%%%%%%%%%%%%%%%%%%%%%
\bibliographystyle{JHEP}
\bibliography{reference.bib}

\end{document}